 \documentclass[12pt,preprint]{aastex}

\shorttitle{Stellar Structure in Triangulum-Andromeda}

\shortauthors{Rocha-Pinto et al.}

\begin{document}


\title{Exploring Halo Substructure with Giant Stars: 
A diffuse star cloud or tidal debris around the Milky Way in Triangulum-Andromeda}

\author{Helio J. Rocha-Pinto, Steven R. Majewski, M. F. Skrutskie, 
    Jeffrey D. Crane, Richard J. Patterson} 
\affil{Department of Astronomy, University of Virginia,
    Charlottesville, VA 22903}
\email{helio, srm4n, mfs4n, jdc2k, rjp0i@virginia.edu}

\begin{abstract}

We report here the discovery of an apparent excess of 2MASS M giant candidates  
with dereddened $ 0.85 < J-K_S < 1.2 $ spanning a considerably large area of the celestial 
sphere between, at least, $100\degr < l < 150\degr$ 
and $-20\degr > b > -40\degr$, and covering most of the constellations of Triangulum and 
Andromeda. 
This structure does not seem to be preferentially 
distributed around a clear core, but rather lies in a tenuous, clumpy cloud-like 
structure tens of kiloparsecs away.  The reduced proper-motion diagram as well as spectroscopy of a 
subsample shows these excess stars to be real giants, not contaminating dwarfs. 
Radial velocity measurements indicate among those M giants the presence of a coherent kinematical structure 
with a velocity dispersion $\sigma < 17$ km s$^{-1}$. 
Our findings support the existence of a quite dispersed stellar structure around the Milky Way that, 
due to its coreless and sparse distribution, could be part of a 
tidal stream or a new kind of satellite galaxy.

\end{abstract}

\keywords{Galaxy: structure -- Galaxy: disk -- galaxies: interactions}

\section{Introduction}

Recent surveys of the Galactic stellar content (e.g.,~\citealt{ibata, ibata2};
\citealt[][ 2003, hereafter M03]{maj96}; \citealt[][ hereafter N02]{newberg}) 
have consolidated the notion 
that galactic mergers play a fundamental role in building our 
Milky Way (MW).  A prominent feature in 
the Galactic halo is the tidal stream of the Sagittarius 
dSph, which circles the MW in a nearly polar orbit 
(\citealt[][ M03]{ibata3}).   
Another companion of our Galaxy remained hidden behind the 
veil of MW disk stars until its debris was 
discovered by N02 as an overdensity of A-F dwarfs in Monoceros. 
By using a statistical method to find relative stellar distances for a
population with unknown [Fe/H], \citet[][ hereafter R03]{R03} found 
M giants from this disrupted satellite galaxy distributed over a large area of sky (see
also M03). 
Given the evidence that the MW halo is highly substructured
\citep[e.g.]{majpadua}, it is likely that additional 
companions of the Milky Way remain to be discovered \citep[see also]{willman}.

In the investigation of the N02 tidal stream, 
R03 pointed out smaller, more distant ($R_{GC} > 20$ kpc) M giant overdensities that
did not seem to be part of that structure. 
We have undertaken a systematic investigation 
of two of these more distant overdensities using a larger subsample of 2MASS M giant candidates. 
Here we show that at least one of these more distant M giants constitutes part of 
a new coherent structure that spans at least $50\degr\times 20\degr$ in the 
second Galactic quadrant, over much of the constellations of Triangulum and Andromeda. 

%In \S 2 of this communication, we discuss the selection criteria defining the stellar sample 
%used in this analysis. In \S 3, stars beyond the Galactic disk are mapped  
%from asymmetries in the 2MASS catalogue along lines-of-sight 
%unaffected by large reddening.  
%One of these asymmetries cannot be accounted for 
%by presently known Galactic satellites. In \S 4, we present radial velocities 
%and metallicities for dozens of
%stars in the fields where stellar overdensities are found. These data, 
%together with the asymmetries found in \S 3, make a case that we have found a new, coherent
%stellar structure 
%SRM: note the change, which is better than "around"
%in the Milky Way halo.
%around the Milky Way. 

\section{Data}

The stellar sample we explore includes all stars from the 2MASS Point Source
Release (2MASS-PSR) having dereddened colors $0.85 < J-K_S < 1.5$, $K_S < 13.0$ and
photometric quality flag `{\tt AAA}' that also meet the color selection criteria 
for M giants candidates in the $(J-K_S, J-H)$ diagram (M03).  
Reddening and extinction in $J$ and 
$K_S$ were calculated by $(A_J,A_{K_S},E_{J-K_S}) = (0.90, 0.25, 0.65)
\alpha E_{B-V}$, where $E_{B-V}$ is from \citet{schlegel}.
$\alpha$ is an attenuation factor that produces better results when using the 
Schlegel et al. $E_{B-V}$ in regions of high extinction.  The adopted value of $\alpha=\case{2}{3}$,
 as well as the selected extinction coefficients
at $J$ and $K_S$,  best reconstructed the un-extinguished distribution of normal giants
in the $(J-K_S, J-H)$ diagram when applied to raw 
2MASS $J$, $H$, and $K_S$ magnitudes in regions of moderate 
extinction ($A_V\sim 3$).

Relative stellar distances were calculated as in R03:  
The distance probability density function (DPDF) for each star is computed by propagating the 
uncertainty of the stellar metallicity, according to an assumed metallicity probability 
distribution function (MPDF). To facilitate comparison with the R03 distance scale, 
we adopt a Gaussian MPDF, with $\langle {\rm [Fe/H]}\rangle = -1.00$ dex and $\sigma = 
0.40$ dex. Hereafter, we will use as an estimate of the distance 
the mode of the DPDF for each star. R03 give an 
approximate equation for the variation of this distance as a function of [Fe/H].

\section{An Excess of Stars in the Outer Galaxy}

The Galactic Equator and $l=0,180^{\degr}$ meridian can be used in the definition of three 
presumed MW symmetries, namely, equatorial, meridional and diagonal, defined so that 
a stellar field limited by the coordinates $\left(\left[l_1:l_2\right],\left[b_1:b_2\right]\right)$ 
has as analogs the fields $\left(\left[l_1:l_2\right],\left[-b_1:-b_2\right]\right)$, 
$\left(\left[-l_1:-l_2\right],\left[b_1:b_2\right]\right)$ and 
$\left(\left[-l_1:-l_2\right],\left[-b_1:-b_2\right]\right)$, respectively. 
We compare these pairs of analogous fields to search for {\it asymmetries} in
stellar density that may represent Galactic substructure.
We discard from the analysis any 1 ${\rm deg}^2$ field if it, or its counterpart, 
has $E_{B-V} > 0.15$. This reddening limit ensures that the differential reddening 
among the target fields is $A_{K_S}\le 0.03$.

\begin{figure}
\epsscale{1.28}
%\plotone{planeforeign.eps}
\plotone{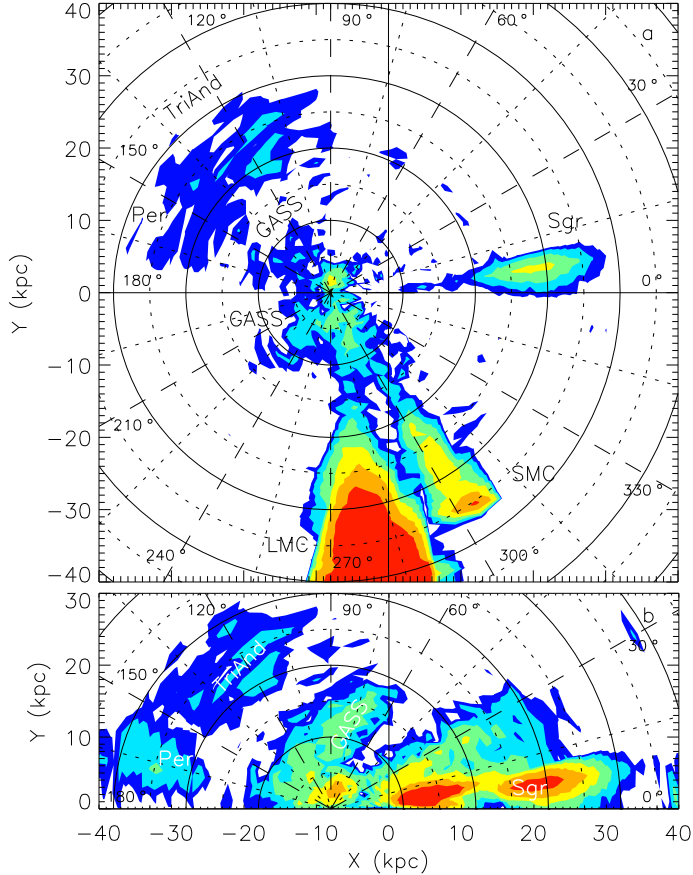}
\caption{Excess stars in the Milky Way vicinity. (a) Polar contour plot 
(circles centered on the Sun)
showing the density of excess stars projected onto the Galactic plane. Contour levels
represent, at $R_{GC} \sim 30$ kpc, 
densities of nearly 10, 20, 40, 80, 160 and 320 stars kpc$^{-2}$.
Most of the 
stellar overdensities correspond to known Milky Way satellites --- the LMC, SMC, and 
Sagittarius galaxies --- or an extension of the recently system first discovered in
Monoceros by N02, labeled GASS (Galactic Anticenter Stellar Stream) in the plot.  
A new, large area overdensity of stars, here called TriAnd, 
can be seen between 15 and 30 kpc from the Sun, between the longitudes $l\sim 100\degr$ 
and $l\sim 170\degr$.  (b) Expanded view of
the TriAnd region when R03's reddening limit of $E_{B-V} <0.55$ is adopted. 
Per was initially found when we limited the sample 
to this reddening value.  With a more conservative reddening limit, Per
is mostly eliminated as an excess feature. 
We opted to include its stars in this discussion to test the possibility 
that Per and TriAnd are associated.
\label{symm}}
\end{figure}

Figure~\ref{symm}a summarizes this symmetry analysis in a composite polar contour plot of 
excess stars in the Galactic neighborhood. The densities correspond to the positive excesses  
in stellar density between opposite hemispheres across each symmetry, 
averaged over the three symmetries considered above.

Several features in the plot can be identified with known satellites of the Milky Way, 
reinforcing our claim that we recover significant halo substructures with this method.
The precise distance range where these structures make their 
appearance in the plot is not accurate for some satellites because distances 
were calculated for a population with $\langle{\rm [Fe/H]}\rangle = -1.0\pm 0.4$.  
Residual photometric errors, metallicity spread and extinction effects artificially 
stretch the distance range covered by
these structures, giving rise to ``Finger of God" effects. 
Pieces of the extensive tidal tail structure that in our previous papers  
we dubbed the Galactic Anticenter Stellar Structure (GASS) 
--- associated with N02's discovery in Monoceros ---
are seen in Figure \ref{symm}a at $110\degr<l<170\degr$ and
$210\degr < l < 270\degr$ with heliocentric distances between 5 and 13 kpc.
In this plot, GASS's signal is not very strong nor does it appear as a continuous stream (as in R03)
due to the severe limit on $E(B-V)$ adopted.

Here we call attention to an additional, unexplained, large grouping of apparent M giants
stretching from $l\sim 100\degr$ to $l\sim 150\degr$ between 15 to 30 kpc from the Sun.  Two 
areas of higher density are in the constellations Triangulum and Andromeda (hereafter, ``TriAnd"). 
Another grouping (hereafter, ``Per"), at $155\degr < l < 174\degr$ and $d\ga 18$ kpc, lies
in Perseus. The latter group is more easily seen in Figure~\ref{symm}b, when we apply a less 
rigorous reddening limit. 
Both panels give the impression that Per is related to TriAnd, but, as shown below, 
they are likely to be unrelated. These same concentrations were already found by R03 as 
secondary peaks within the DPDF at the corresponding lines of sight.

\begin{figure}
\epsscale{1.3}
\plotone{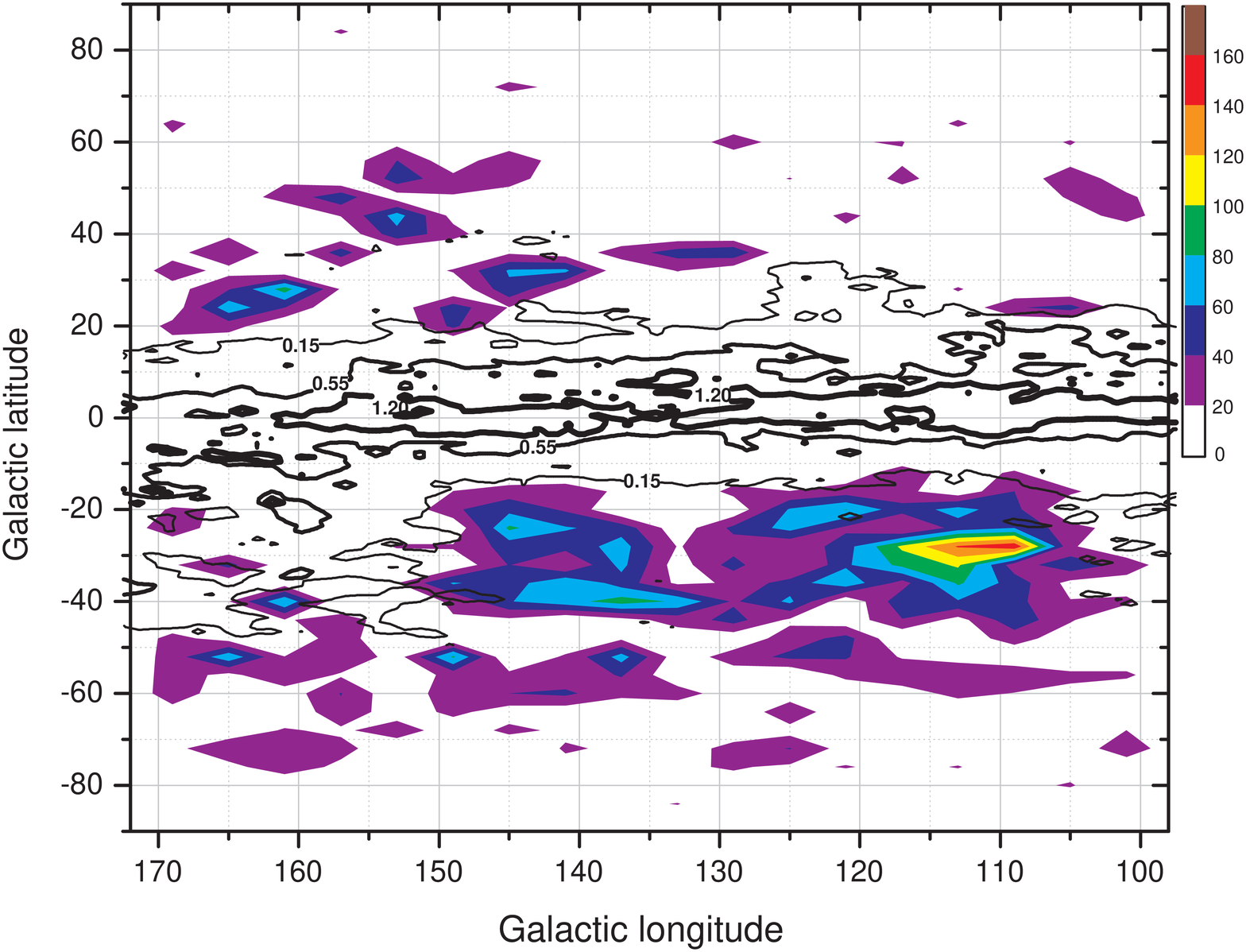}
\caption{2MASS M giant candidate overdensity having $0.85 < J - K_S \le 1.50$, 
$K_S < 13.0$, $15 < d < 40$ kpc and $E_{B-V} < 0.15$ in the region of Triangulum and 
Andromeda, averaged over the three Galactic symmetries.
Unshaded contour levels show the average reddening 
in this sky field at $E_{B-V}=0.15$, 0.55 and 1.20. The other  
concentration considered in this work, in Perseus, centered at $(l,b)=(163\degr,-20\degr)$, 
is not shown in this plot, due to the conservative reddening limited adopted. 
\label{skyview}}
\end{figure}

Figure~\ref{skyview} presents an on-sky projection of the residual 
density of M giant candidates having $15 < d < 40$ kpc in the TriAnd area, 
averaged over the three Galactic symmetries. The figure shows the clear 
asymmetry in the M giant distribution across the Galactic plane, however, 
as in Figure \ref{symm}, this projection shows no obvious core as would be expected
for a typical dwarf galaxy. 
The presence of considerable foreground dust may hinder the search for a core in this 
region, as can be 
appreciated from the large open areas that we have been forced to ignore due to 
excessive reddening.

\begin{figure}
\epsscale{0.83}
\plotone{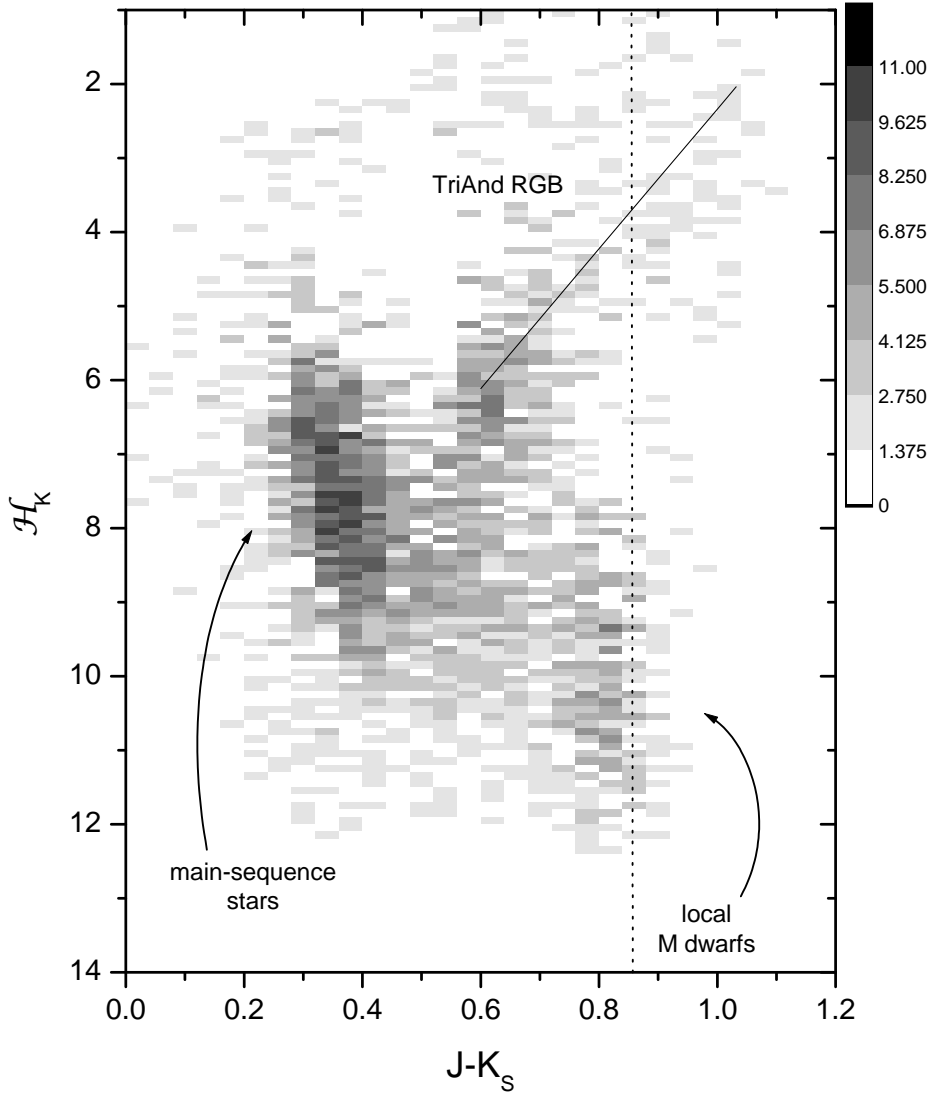}
\caption{Hess-like diagram showing the residual RPMD of stars in 
the TriAnd field with $11.0 < K_S < 12.5$ after subtracting the field with diagonal 
Galactic symmetry and discarding stars in all 1 deg$^2$ bins (and symmetric counterparts) 
reddened by $E_{B-V} > 0.15$. The color bar scales to $\sqrt{N}$, where $N$ is the number 
of stars in each diagram cell. The dotted vertical line indicate the original blue color limit 
used in the discovery of TriAnd and in Figures 1 and 2. 
A solid line marks the expected locus of the RGB for a population 
with ${\rm [Fe/H]} = -1.0$ dex, according to \citet{ivanov}. 
%A strong blue feature with $J-K_S \la 0.40$ in 
%this plot is caused by a group of local turnoff main-sequence stars.
\label{rpmd}}
\end{figure}

We have considered the possibility that TriAnd is an artifact created by 
misclassified M dwarfs. The separation between M giants and dwarfs in 
the $(J-K_S, J-H)$ diagram begins at 
$J-K_S \sim 0.85$ \citep{bessell}, so we expect some interloping 
dwarfs at the blue limit of our sample, due 
to photometric errors and cosmic scatter. 
Indeed, we have found in our data a kind of shell around the Sun, along almost every line of 
sight, composed of stars having $0.85 < J-K_S < 0.90$ and $12.5 < K_S < 13.0$, which we 
take to be misclassified M dwarfs (see discussion of a similar 
``contamination shell" effect in M03). The 
apparent magnitudes of these shell interlopers are  
translated to similar stellar distances as redder, intrinsically
brighter TriAnd M giants.
While a {\it symmetric} shell of such M dwarf interlopers should not contribute to our ``excesses"
in Figures \ref{symm} and \ref{skyview}, if sufficiently dominant, statistical fluctuations might conspire 
to create a false excess signal, as could {\it actual} local M dwarf substructure.
To assess the influence of M dwarf interlopers, we use a reduced proper-motion diagram 
(hereafer, RPMD) for luminosity discrimination. 
A `proper-motion' estimate, $\mu$, was calculated from the 2MASS-PSR 
parameter {\tt dist\_op}, which gives the displacement between the 
2MASS source position 
and an associated optical source in some published catalogues.
%either the Tycho 2 or USNO-A2.0 catalogue. 
%We discard 
%all stars for which {\tt dist\_op} was calculated with respect to Tycho 2, 
%because these displacements are not likely to give reliable $\mu$ due to the 
%short time baseline to those observations. 
We used only {\tt dist\_op} measured with respect to USNO-A2.0. This parameter  
was divided by 50 or 20 yr, for 
declinations above or below $-30\degr$, respectively, to take into account the 
approximate time baseline from the first epoch observations made at Palomar and ESO, 
respectively. A TriAnd target field  was defined by the range 
$(l,b) = \left(\left[105\degr:125\degr\right],\left[-18\degr:-40\degr\right]\right)$, 
corresponding  to the area where TriAnd appears strongest.  We extract from the 
2MASS catalogue all stars having $11.0 < K_S < 12.5$ in the TriAnd and symmetric fields
(since this is the magnitude range that has been translated to the TriAnd distances in Fig. 
\ref{symm}) and take the difference between the averaged RPMD of the 
symmetric fields and the TriAnd field RPMD. An example of this extracted RPMD is shown in 
Figure~\ref{rpmd}.  We have used diagonal Galactic symmetry to construct this 
particular plot; the other two symmetry tests yield similar results. 
The RPMD quantity ${\cal H}_K$ is given by ${\cal H}_K \equiv K_S + 5\log\mu +5$.  
Figure \ref{rpmd} shows a residual red giant branch (RGB) down to a red clump at 
$J-K\sim 0.6$ in the 
TriAnd RPDM.  The theoretical RGB isochrone for ${\rm [Fe/H]} = -1.0$ \citep{ivanov}
gives a reasonable fit to the observed RGB. 
%when shifted by $8.5\pm 0.3$ mag, which corresponds to 
%$|T|\approx 230\pm 40$ km/s, according to Equation~\ref{reduced}. 
Nevertheless, figure \ref{rpmd} suggests that some M dwarfs with 
$J-K_S \gtrsim 0.85$ interlope in the sample, but the number of TriAnd M giants greatly 
increases with this color selection.

 Another feature visible in the RPMD is 
a significant excess of main sequence turnoff stars with $J-K_S\la 0.40$; the corresponding
mean apparent magnitude of these stars is $K_S\approx 12.1$.  The turnoff 
and M dwarfs are local features that should be located at 500-1000 pc and 150-200 pc, according 
to typical $JHK_S$ magnitudes for a G0 and M4 dwarf, respectively. Therefore, these features
are unrelated to the RGB stars with $J-K_S >  0.90$, since, with
$K_S > 11.0$, such RGB stars are located much farther away from the Sun, presuming 
an average absolute magnitude of $\langle M_{K_S} \rangle \sim -5.5\pm 1.1$ (Rocha-Pinto et al. 2004, 
in preparation).

\section{Spectroscopy of Triand Stars}

A test of the nature of the TriAnd excess comes via the kinematical properties
of its constituent stars. 
We collected spectra for 36 TriAnd and ten Per M giants, as well 
as 51 spectra of twenty radial velocity standard stars on UT 2003 
December 13--18, as part of an ongoing M giant radial velocity program using the B\&C 
Spectrograph on the Bok 2.3 m telescope at Steward Observatory. The spectra cover 
7800 \AA\ $< \lambda <$ 9000 \AA\ at $R \sim 3300$. 
Radial velocity standards were taken from the IAU standard list and from stars in
\citet{bettoni}. 
Spectral reduction and $v_r$ calculations followed the procedure outlined by 
\citet{crane}. Our internal radial velocity accuracy ($\sim$ 8 km s$^{-1}$ among  
$v_r$ standards) is larger than 
the 2.7 km s$^{-1}$ achieved by \citet*{crane} due to differences in resolution and a slight
undersampling of the Bok data compared to Crane et al.'s spectra.

\begin{figure}
\epsscale{1.0}
\plotone{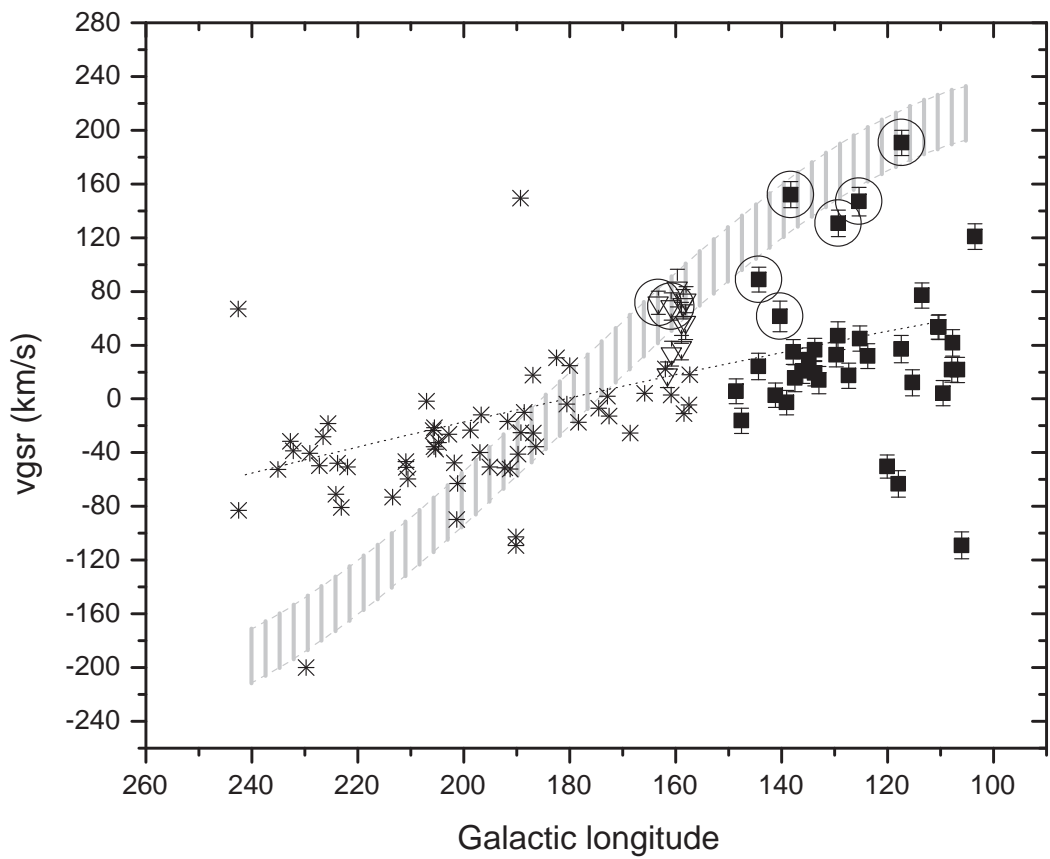}
\caption{Radial velocity of program stars in the Galactocentric Standard of Rest 
as a function of Galactic longitude. Stars 
in the TriAnd and Per groups are shown by squares and upside-down triangles, respectively. 
We also show velocities for GASS stars published by \citet[][ asterisks]{crane}. 
The hatched area is the expected 2$\sigma$ region for local M dwarfs following a circular orbit
with rotation velocity $\Theta_{\rm LSR}$, $\sigma=10$ km s$^{-1}$, about the Galactic Center, 
while the dotted curve is for
a similar rotational velocity at $R_{\rm GC} = 30 kpc$. Stars classified as dwarfs 
or having undefined classification, according to the presence of strong Na I doublet 
lines the spectrum, are marked by an open circle.
\label{radvel}}
\end{figure}

Our observations are summarized in Table 1, where we give, for each star, respectively, 
its name, Galactic coordinates, 
Julian day of observation, estimated S/N ratio per pixel at $\lambda\sim  8630$ \AA, 
$K_S$, $J-K_S$, the mode of the DPDF, heliocentric $v_r$ and its associated error,
 derived metallicity (see below) and the overdensity to which the star belongs, respectively. 
Figure~\ref{radvel} summarizes our $v_r$ measurements of TriAnd and Per stars, transformed to the Galactic 
Standard of Rest, using $\Theta_{\rm LSR}=220$ km s$^{-1}$. 
Only a few TriAnd stars have $v_r$ consistent with being 
disk M dwarfs.  Moreover, most TriAnd stars show a 
relatively small observed velocity dispersion [$\sigma 
\sim 19.5$ km/s around a linear fit of the $v_r(l)$ trend (not shown in the plot), 
after discarding stars deviating by more than 3$\sigma$] and have an 
average velocity ($\sim 27$ km s$^{-1}$) inconsistent with 
a circular orbit at $R_{\rm GC}= 30$ kpc assuming a flat Galactic rotation curve. 
In contrast, Perseus stars have $\langle{v_r}\rangle=57$ km/s --- i.e. 
not completely inconsistent with being disk stars; their observed dispersion is 
$\sigma\sim 20.7$ km s$^{-1}$.
Taking into account an average observed error in $v_r$ for these stars of 10 km s$^{-1}$,
the real velocity dispersion of the TriAnd and Per samples are 17 and 18 km s$^{-1}$, 
respectively.  The $v_r(l)$ trend found among TriAnd stars does not  seem to connect with 
the Perseus stars, despite our initial expectation that they were part of the same structure. 
On the other hand, the TriAnd velocity trend {\it does} fall along that of GASS, 
suggesting a possible connection, though 
we stress that the two features are at different distances, with GASS {\it interior} to TriAnd
\citep[][ in preparation]{maj04}\footnote{We have considered the possibility that the TriAnd 
excess is a false signal due to improper photometric distances assigned to GASS stars. 
Such stars would need to have an [Fe/H] exceeding +1.0 dex to reduce their distances to that 
of the foreground GASS stars. But these metallicities are strongly ruled out by the 
[Fe/H] values calculated for some TriAnd stars in this paper.}.  

The NIR Na I doublet ($\lambda\lambda 8193,8195$) is a well known index sensitive 
to $\log g$ and can be used to classify the stars we observed as giants or dwarfs 
\citep{schiavon}. Six of the stars in Table 1 have red dwarf gravities and two stars have 
undefined spectra, although they are more likely to be dwarfs than giants. The other stars 
have red giant spectra, according to the standards given by 
\citet{schiavon}. The red dwarfs are just those stars in the TriAnd longitude range that 
have $v_r$ closer to the expected local trend in Figure \ref{radvel}. The last column of Table 1 
indicates whether the star have dwarf (`D'), giant (`G') or undefined (`U') spectra.

Metallicities were calculated from the same spectra
by comparing the sum of three Ca IRT spectral indices for 
target and standard stars, as explained by \citet{crane}. Only spectra with a $S/N$ higher 
than 30 were used in this calculation. The result is an average metallicity ${\rm [Fe/H]} 
= -1.2$, with $\sigma\sim 0.5$ dex, for twenty TriAnd stars.
Per stars are more metal poor, with $\langle{\rm [Fe/H]}\rangle \le -1.5$. 
Metallicity was not calculated for the stars spectroscopically identified 
as dwarfs, since our relation between [Fe/H] and the Ca IRT indices was calibrated 
with M giant standards.

\section{Discussion} 

A remarkable aspect of the TriAnd feature is that it is very diffuse and large.
This no doubt conspired to keep it hidden among Milky Way stars, in spite of 
it being at relatively high ($l\sim30^{\circ}$) Galactic latitude.
In a companion paper, \citet{maj04} detect the more populous, though
still diffuse TriAnd main
sequence turn-off, and from the rather constant areal density of these stars calculate a luminous
mass of $\sim 1.6\times 10^6 M_\odot$. These are properties unlike any known
Milky Way satellite.  If a bound system, TriAnd must be a very massive, but very {\it dark}
satellite.  Our calculated velocity dispersion 
would be consistent with a system having a very large $M/L$.  

Alternatively, with such a low luminous mass, TriAnd could be 
a largely unbound cloud of stars from a completely disrupted dwarf galaxy.  A 
recent catastrophic  disruption could explain a diffuse, apparently coreless but still
round configuration, and why the system has not yet stretched into tidal tails 
--- although, such tidal tails would be even {\it more} diffuse than the center of the
system, and therefore even more difficult to detect.  On the other hand, TriAnd
may itself {\it be} tidal debris, perhaps just one higher density part of a longer, 
but diffuse tidal tail system. 
%The ``Finger of God" effects observed for other satellites in Figure \ref{symm} may artificially
%stretch Triand in the radial dimension, suggesting that it is possibly thinner than
%it looks in that projection. 

An obvious possibility is that TriAnd is an old piece of the 
system creating the closer GASS tail, perhaps a different, larger radius orbital wrap of this satellite 
galaxy.  A similar orbital shape might explain the apparent continuous velocity
trend from GASS to TriAnd stars seen in 
Figure~\ref{radvel}. The more metal poor content of TriAnd M giants compared to GASS
 M giants 
\citep{crane} would be consistent with the idea that TriAnd is older both in terms of 
stellar ages and dynamical ages than the inner GASS stars.  We plan to address these 
questions as well as the nature of Per in an upcoming paper.

\acknowledgements

We acknowledge funding by NSF grant AST-0307851, NASA/JPL contract 1228235, the David and 
Lucile Packard Foundation, and the 
generous support from Frank Levinson and Wynnette LaBrosse 
through the Celerity Foundation.

%\begin{figure}
%\epsscale{.7}
%\plotone{f5.eps}
%\caption{Near-infrared theoretical isochrones calculated by \citet{girardi} 
%for solar composition dwarfs compared to giants 
%having a range of metallicities. A dashed vertical line at $K_S = 0.85$ shows the empirical color 
%where the separation of dwarfs and giants begin in this diagram. The several curves for giants 
%indicate that this separation does not depend on the metallicity of the giants.
%\label{f5}}
%\end{figure}

\begin{deluxetable}{cccccrccrrlcc}
\rotate
\tabletypesize{\scriptsize}
\tablecaption{Data for program stars.}
\tablewidth{0pt}
\tablehead{
\colhead{name}  & \colhead{$l$} & \colhead{$b$} & \colhead{JD} & 
\colhead{S/N} & \colhead{$K_S$} & \colhead{$J-K_S$} & \colhead{$d$ (kpc)} & \colhead{$v_r$ (km/s)} 
& \colhead{$\epsilon_{v_r}$} & \colhead{[Fe/H]} & \colhead{group} & \colhead{Spec.}
}
\startdata
2MASSX J00091797+3630588 & 113.56 & $-$25.59 & 2452987.678 & 25 &  9.86 & 1.20 & 25.06 &  $-$110.7 &     9.6 &  $-$1.2 & TriAnd & G \\
2MASSX J00191095+3348239 & 115.30 & $-$28.59 & 2452987.688 & 60 & 10.06 & 1.17 & 25.28 &  $-$167.3 &     9.7 &  $-$1.0 & TriAnd & G \\
2MASSX J00253106+3828397 & 117.38 & $-$24.11 & 2452989.630 & 35 & 12.08 & 0.92 & 25.51 &      +8.0 &     9.6 &  \nodata & TriAnd & D \\
2MASSX J00273574+3504151 & 117.43 & $-$27.55 & 2452990.631 & 25 & 12.04 & 0.96 & 29.69 &  $-$137.9 &    10.2 &  $-$1.3 & TriAnd & G \\
2MASSX J00282905+3753390 & 117.95 & $-$24.76 & 2452990.621 & 40 & 11.41 & 1.03 & 28.04 &  $-$245.1 &    10.0 &  $-$0.8 & TriAnd & G \\
2MASSX J00373441+4042425 & 120.10 & $-$22.08 & 2452989.643 & 35 & 11.42 & 1.01 & 26.48 &  $-$210.4 &     8.5 &  $-$0.8 & TriAnd & G \\
2MASSX J00545179+3510599 & 123.72 & $-$27.68 & 2452990.643 & 35 & 11.50 & 0.98 & 24.46 &  $-$130.9 &     9.4 &  $-$1.1 & TriAnd & G \\
2MASSX J01002299+2957238 & 125.24 & $-$32.88 & 2452987.697 & 50 &  9.72 & 1.18 & 22.30 &  $-$107.7 &     9.6 &  $-$0.8 & TriAnd & G \\
2MASSX J01023690+3735148 & 125.38 & $-$25.23 & 2452991.629 & 40 & 11.91 & 0.94 & 25.28 &   $-$17.1 &    10.7 &  \nodata & TriAnd & D \\
2MASSX J01110387+3553193 & 127.39 & $-$26.81 & 2452987.704 & 30 &  9.28 & 1.27 & 24.69 &  $-$140.5 &     9.5 & $<-1.5$ & TriAnd & G \\
2MASSX J01172688+3218515 & 129.29 & $-$30.24 & 2452990.653 & 25 & 12.03 & 0.97 & 30.38 &   $-$14.3 &     9.8 &  \nodata & TriAnd & U \\
2MASSX J01214158+3635505 & 129.68 & $-$25.88 & 2452987.709 & 50 & 10.65 & 1.06 & 22.53 &  $-$121.5 &     9.1 &  $-$0.9 & TriAnd & G \\
2MASSX J01234407+4131080 & 129.40 & $-$20.95 & 2452989.654 & 30 & 11.77 & 0.94 & 24.32 &  $-$116.5 &     9.6 &  $-$0.6 & TriAnd & G \\
2MASSX J01372969+3719375 & 132.99 & $-$24.64 & 2452989.672 & 35 & 11.29 & 1.00 & 23.50 &  $-$133.2 &    10.3 &  $-$1.2 & TriAnd & G \\
2MASSX J01403652+3649067 & 133.77 & $-$25.02 & 2452987.717 & 25 & 11.08 & 1.02 & 23.58 &  $-$107.0 &     8.6 &  $-$0.1 & TriAnd & G \\
2MASSX J01425641+3851201 & 133.79 & $-$22.93 & 2452987.725 & 30 & 10.73 & 1.07 & 23.81 &  $-$127.9 &     9.4 &  $-$0.8 & TriAnd & G \\
2MASSX J01450678+3623258 & 134.85 & $-$25.24 & 2452987.732 & 20 & 10.65 & 1.17 & 32.87 &  $-$105.3 &     9.3 &  $-$1.1 & TriAnd & G \\
2MASSX J01513509+2339258 & 140.29 & $-$37.18 & 2452990.666 & 20 & 12.24 & 0.94 & 29.69 &   $-$43.3 &    11.2 & \nodata & TriAnd & D \\
2MASSX J01561741+4044275 & 135.98 & $-$20.48 & 2452989.662 & 30 & 11.47 & 1.00 & 26.33 &  $-$122.5 &     9.5 &  $-$1.4 & TriAnd & G \\
2MASSX J01581477+1827407 & 144.32 & $-$41.62 & 2452991.644 & 25 & 12.07 & 0.92 & 25.97 &    $-$5.7 &     9.6 & \nodata & TriAnd & D \\
2MASSX J02001321+3547225 & 138.28 & $-$25.02 & 2452987.739 & 30 & 10.99 & 1.08 & 28.18 &     +21.0 &     9.7 & \nodata & TriAnd & D \\
2MASSX J02044137+4059528 & 137.53 & $-$19.79 & 2452987.748 & 40 & 11.11 & 1.00 & 22.39 &  $-$124.5 &    10.1 &  $-$0.7 & TriAnd & G \\
2MASSX J02075025+4201519 & 137.79 & $-$18.63 & 2452987.758 & 30 &  9.92 & 1.20 & 25.97 &  $-$105.1 &     9.3 & $<-1.5$ & TriAnd & G \\
2MASSX J02133023+4131328 & 139.02 & $-$18.77 & 2452989.685 & 30 & 11.24 & 1.05 & 27.81 &  $-$139.1 &     9.2 & $<-1.5$ & TriAnd & G \\
2MASSX J02195064+3850488 & 141.21 & $-$20.87 & 2452991.658 & 25 & 11.95 & 0.93 & 25.51 &  $-$125.7 &     9.0 &  $-$0.2 & TriAnd & G \\
2MASSX J02222634+3216028 & 144.42 & $-$26.78 & 2452990.712 & 30 & 11.41 & 1.00 & 24.91 &   $-$77.4 &    10.5 &  $-$1.2 & TriAnd & G \\
2MASSX J02481130+3650021 & 147.54 & $-$20.40 & 2452989.695 & 25 &  9.81 & 1.19 & 23.86 &  $-$124.5 &     9.5 & $<-1.5$ & TriAnd & G \\
2MASSX J02541392+3702586 & 148.57 & $-$19.64 & 2452990.725 & 18 & 11.80 & 0.94 & 24.41 &   $-$98.6 &     9.2 &  $-$0.8 & TriAnd & G \\
2MASSX J03244829+3030073 & 158.06 & $-$21.72 & 2452989.701 & 15 & 10.48 & 1.12 & 25.65 &      +6.5 &     9.4 &  $-$1.2 &  Per  & G \\
2MASSX J03245531+3026213 & 158.12 & $-$21.75 & 2452989.712 & 20 & 10.73 & 1.06 & 23.13 &    $-$9.3 &    13.2 & $<-1.5$ &  Per  & G \\
2MASSX J03283560+3042261 & 158.62 & $-$21.06 & 2452991.701 & 15 & 11.49 & 1.02 & 28.18 &      +4.0 &    11.1 &  $-$1.0 &  Per  & G \\
2MASSX J03290508+3022080 & 158.93 & $-$21.27 & 2452989.730 & 50 &  8.29 & 1.45 & 29.85 &   $-$29.3 &     9.2 &  $-$1.1 &  Per  & G \\
2MASSX J03303497+3050191 & 158.90 & $-$20.70 & 2452990.737 & 12 & 11.62 & 0.99 & 26.48 &    $-$4.0 &    14.2 & $<-1.5$ &  Per  & G \\
2MASSX J03355851+3107296 & 159.69 & $-$19.76 & 2452991.715 & 15 & 11.42 & 0.97 & 23.21 &     +15.2 &    12.3 & $<-1.5$ &  Per  & G \\
2MASSX J03415347+2954439 & 161.55 & $-$19.90 & 2452987.766 & 30 & 10.34 & 1.15 & 26.84 &   $-$40.4 &     9.7 &  $-$1.4 &  Per  & G \\
2MASSX J03443920+3144508 & 160.79 & $-$18.11 & 2452987.775 & 25 & 10.13 & 1.16 & 25.28 &   $-$26.4 &     9.5 &  $-$1.4 &  Per  & G \\
2MASSX J03453551+3156257 & 160.82 & $-$17.83 & 2452989.741 & 20 & 11.28 & 1.01 & 24.55 &      +6.9 &    10.0 & \nodata &  Per  & U \\
2MASSX J04021237+3209591 & 163.41 & $-$15.33 & 2452989.755 & 20 & 11.96 & 0.93 & 25.14 &   \nodata & \nodata & $<-1.5$ &  Per  & G \\
2MASSX J04043600+3343381 & 162.68 & $-$13.85 & 2452991.856 & 20 & 11.91 & 0.94 & 25.74 &   \nodata & \nodata & $<-1.5$ &  Per  & G \\
2MASSX J04063385+3316474 & 163.31 & $-$13.89 & 2452991.871 & 30 & 11.42 & 0.96 & 22.30 &     +14.0 &    10.6 & \nodata &  Per  & D \\
2MASSX J23280198+3312451 & 103.50 & $-$26.49 & 2452987.662 & 20 & 11.24 & 1.02 & 25.51 &   $-$79.1 &     9.8 &  $-$0.6 & TriAnd & G \\
2MASSX J23333834+3909235 & 106.85 & $-$21.27 & 2452987.639 & 25 & 10.60 & 1.07 & 22.62 &  $-$180.8 &     9.5 &  $-$0.1 & TriAnd & G \\
2MASSX J23382512+3927458 & 107.89 & $-$21.27 & 2452990.587 & 30 & 11.64 & 0.98 & 26.39 &  $-$181.6 &     9.4 &  $-$1.4 & TriAnd & G \\
2MASSX J23413614+3143111 & 105.99 & $-$28.84 & 2452990.600 & 40 & 11.17 & 1.04 & 26.48 &  $-$300.2 &    10.0 &  $-$1.4 & TriAnd & G \\
2MASSX J23485366+3144486 & 107.70 & $-$29.27 & 2452990.610 & 30 & 11.30 & 1.04 & 27.89 &  $-$147.6 &     9.8 &  $-$0.9 & TriAnd & G \\
2MASSX J23490540+4057312 & 110.43 & $-$20.39 & 2452989.603 & 25 & 11.11 & 1.02 & 23.35 &  $-$145.4 &     9.1 &  $-$1.1 & TriAnd & G \\
2MASSX J23534927+3659173 & 110.32 & $-$24.47 & 2452989.612 & 25 & 11.02 & 1.07 & 26.84 &  $-$142.3 &     9.4 &  $-$1.4 & TriAnd & G \\
2MASSX J23583475+3009356 & 109.55 & $-$31.34 & 2452989.623 & 45 & 10.08 & 1.18 & 26.33 &  $-$178.5 &     9.3 & $<-1.5$ & TriAnd & G \\

\enddata
%% You can append references to a table using the \tablerefs command.
\end{deluxetable}

\end{document}